\documentclass[12pt,a4]{article}
\usepackage{graphicx} 
\usepackage{amsmath}
\usepackage{amssymb}
\usepackage[margin=1in]{geometry}
\usepackage{cite}
\usepackage{hyperref}
\usepackage{cleveref}
\usepackage{float}
\usepackage{scalerel}
\usepackage[symbol]{footmisc}

\usepackage[utf8]{inputenc}
\usepackage{soul}
\usepackage[dvipsnames]{xcolor}
\usepackage{lineno}
\definecolor{orchid}{HTML}{AA00DD}

\newcommand*{\Scale}[2][4]{\scalebox{#1}{$#2$}}%

\graphicspath{figures/}

\begin{document}

\begin{center}
    {\Large
    \textbf{Spin 1 Transverse Momentum Dependent Tensor Structure Functions in CLAS12 }
    }\\
    \vspace{5pt}
    {\large
    CLAS12 Analysis Proposal
    }
    \vspace{15pt}

    I.~P.~Fernando, D.~Keller \\
    \emph{University of Virginia, VA}
    
    \vspace{11pt}

    E.~Long, D.~Ruth, K.~Slifer, S.~N.~Santiesteban\\
    \emph{University of New Hampshire}
    \vspace{11pt}

    A. Bacchetta \\ 
    \emph{University of Pavia, IT and INFN Pavia, IT}
    \vspace{11pt}
    
    J.~P.~Chen, J.~Poudel\footnote[2]{Contact: jpoudel@jlab.org}\\
    \emph{Thomas Jefferson National Accelerator Facility, VA}
    \vspace{11pt}

    \date{}
    
\end{center}

\begin{center}
    \textbf{Abstract}\\ 
\end{center}

\noindent We propose to analyze CLAS12 RG-C data to study the tensor transverse-momentum-dependent parton distribution functions (TMDs) on deuteron data. The deuteron is the lightest nucleus with spin-1, in essence a weakly bound system of two spin-1/2 nucleons. However, one of the most intriguing characteristics of the deuteron is that the tensor polarized structure provides direct access to the quark and gluon distribution of light nuclear system, which cannot be naively constructed from the proton and neutron. We will study the tensor polarized structure functions with the Semi-inclusive Deep Inelastic Scattering (SIDIS) $eD \rightarrow e'P_{h}X$ and Inclusive processes in the available data on deuterated ammonia (ND$_3$) target. We will perform the first ever SIDIS analysis extraction of the tensor structure functions, which can be interpreted in term of completely unexplored tensor polarized TMDs. 

Our analysis will focus on the extraction of the tensor structure functions $b_1$ from inclusive process, and $F_{U(LL),T}$ and $F_{U(LL)}^{\cos{2\phi_h}}$ from SIDIS. These last two structure functions carry information related to two tensor-polarized TMDs, $f_{1LL}$ and $h_{1LL}^\perp$. These initial exploratory measurements of tensor-polarized structure functions will enable the first extraction of spin-1 TMDs and motivate more precise future measurements. 

\clearpage

\section{Introduction}

TMDs describe the 3-D distributions of partons in momentum space and opens new perspective to study Quantum Chromodynamics (QCD) \cite{2023tmdhandbook,bacchetta2016TMD}. In addition to the dependence on the longitudinal momentum fraction $x$, the TMDs also describe the parton distributions as a function of the partonic transverse momentum $k_T$.
While the spin-$\frac{1}{2}$ nature of the nucleons has been studied in previous TMD experiments, the tensor component of spin-1 systems such as light nuclei has been hardly explored. The tensor structure of the spin-1 deuteron will unlock new and fascinating possibilities for further understanding the parton structure in light nuclei.

Theoretical interest in the distribution functions that describe the spin-one targets has increased over the years. Pioneering work by Hoodbhoy, Jaffe, and Manohar described the structure functions of a spin-one target, in particular the leading twist contributions, that could be measured with an unpolarized beam and provide a clear signature for exotic components in a spin-one nucleus~\cite{Hoodbhoy:1988am}.  The simplest spin-one target is the deuteron, which is a weakly bound system of two spin-$\frac{1}{2}$ hadrons. Deuteron targets have historically been used primarily to extract neutron distribution functions in DIS, but there is growing interest in expanding their use to understand the full spin structure of the bound system. The spin-one structure of the deuteron in itself is the subject of a number of interesting questions ~\cite{Nikolaev:1996jy,Bora:1997pi,Umnikov:1996qv,Edelmann:1997qe}.

Tensor structure is not understood fully at the parton level, which suggests that a new field of high energy spin physics could be created by investigating tensor-polarized structure functions. These structure functions enable additional studies of the interaction between partons beyond what can be determined with spin-$\frac{1}{2}$ nucleon structure functions alone \cite{PhysRevD.82.2010_Kumano,proceeding_Kumano_2022}. The spin-1 deuteron has been described as a simple bound system of a proton and a neutron, but the tensor nature of the spin-1 system is missing completely in spin-$\frac{1}{2}$ particles. So, tensor polarized structure functions could lead us to a new hadronic physics providing new information on QCD and nuclear structure.

There have been only few experimental studies of tensor structure functions for spin-one targets due to the experimental challenge of enhancing the tensor polarization in a polarized target. The only measurement to date was a measurement of the $b_1$ tensor structure function by the HERMES Collaboration~\cite{PhysRevLett.95.2005_Hermes}, published almost two decades ago in 2005. There are two leading-twist collinear structure functions in lepton scattering from the deuteron:  $b_1$ and $b_2$~\cite{Frankfurt:1983qs}.  They are related to each other by a Callan-Gross type relation $2xb_1 = b_2$ in the scaling limit on $Q^2$. A phenomenological parametrization of the HERMES data was performed by Kumano in Ref.~\cite{PhysRevD.82.2010_Kumano}. The results from HERMES showed a noticeable difference from conventional convolution calculations based on a standard deuteron model with D-state admixture~\cite{Cosyn:2017fbo,Cosyn:2020kwu}. To study further, a new experiment was approved to measure $b_1$ in Hall C of Jefferson lab via the tensor asymmetry ($A_{zz}$)~\cite{proposal_b1_jlab}. The main difference between the target of the existing CLAS12 data and the proposed tensor experiments at Hall C~\cite{proposal_b1_jlab} is the technique used for the enhancement of tensor polarization. The Hall C experiment will use Dynamic Nuclear Polarization (DNP), Adiabatic Fast Passage and semi-saturating radiofrequency (ss-RF)~\cite{Keller:2020wan,Clement:2023eun} to enhance the tensor polarization to an expected average of about 30\%. On the other hand, RG-C employed only DNP to produce an average vector (or spin-$\frac{1}{2}$) polarization ($S_\parallel$) of about 29\%, which created a corresponding tensor polarization ($T_{\parallel\parallel}$) of about 10\%. 

 Given the clear importance of studying tensor-polarized structure function to shed light on the partonic structure of the nucleus, we plan to analyze the CLAS12 RG-C data to extract several tensor structure functions. RG-C has the deuteron vector and tensor polarization data already on tape and the data set is unique because of the large acceptance of CLAS12. 
 The result of this analysis is crucial to motivate and guide new measurements of these quantities and will lay the foundation for a dedicated tensor TMD program at Jefferson Lab. The analysis will also be critical to develop tools necessary for the analysis of the already approved tensor experiments in Hall C.  


\section{Physics Motivation}
\label{section:motivation}
A complete description of the partonic orbital angular momentum requires an in-depth knowledge of the three-dimensional structure functions. TMDs represent the parton distributions as functions of the the longitudinal momentum fraction $x$ and partonic transverse momentum $k_T$. At twist-2, the tensor component of the spin-1 system adds ten new TMDs accessible only with considerably more complex target configurations as described in Ref.~\cite{PhysRevD.62.2000_bacchetta}. In the CLAS12 data that we will analyze, the target has tensor polarization along the longitudinal axis (LL), which is the simplest case from the theory perspective. The full calculation of the cross section was recently derived in Ref.~\cite{bacchetta:2023Contact}. 
In the following, we present the formalism of the SIDIS cross section for the LL case and the previous measurements of the tensor structure functions. 

\subsection{Formalism}
Consider the single photon exchange in the electron-deuteron ($eD$) interaction
\begin{align}
    e(l)+d(P_d)\rightarrow e(l')+h(P_h)+X
\end{align}
where $l, P_d, l', P_h$ are the four momenta of the incident electron, deuteron target, scattered electron and produced hadron ($h$), respectively. If $M_d~\text{and}~M_h$ represent the masses of deuteron target and outgoing hadron respectively, the various kinematic variables are expressed in terms of four momenta as:
\begin{align}
    x_d=& \frac{Q^2}{2P_d\cdot q} \qquad 0 < x_d < 1\\
    y=& \frac{P_d\cdot q}{P_d\cdot l}\\
    z=& \frac{P_d\cdot P_h}{P_d\cdot q}\\
    \gamma=& \frac{2M_dx}{Q}
\end{align}
where $q=l-l'$ and $Q^2=-q^2 >0$. The conventional scaling variable $x_d$ for deuteron can be expressed as effective Bjorken variable ($x$) for scattering from a nucleon as $x=2x_d$ with $0<x<2$.

\begin{figure}[htbp]
    \centering
    \includegraphics[height=6.5cm,width=12cm]{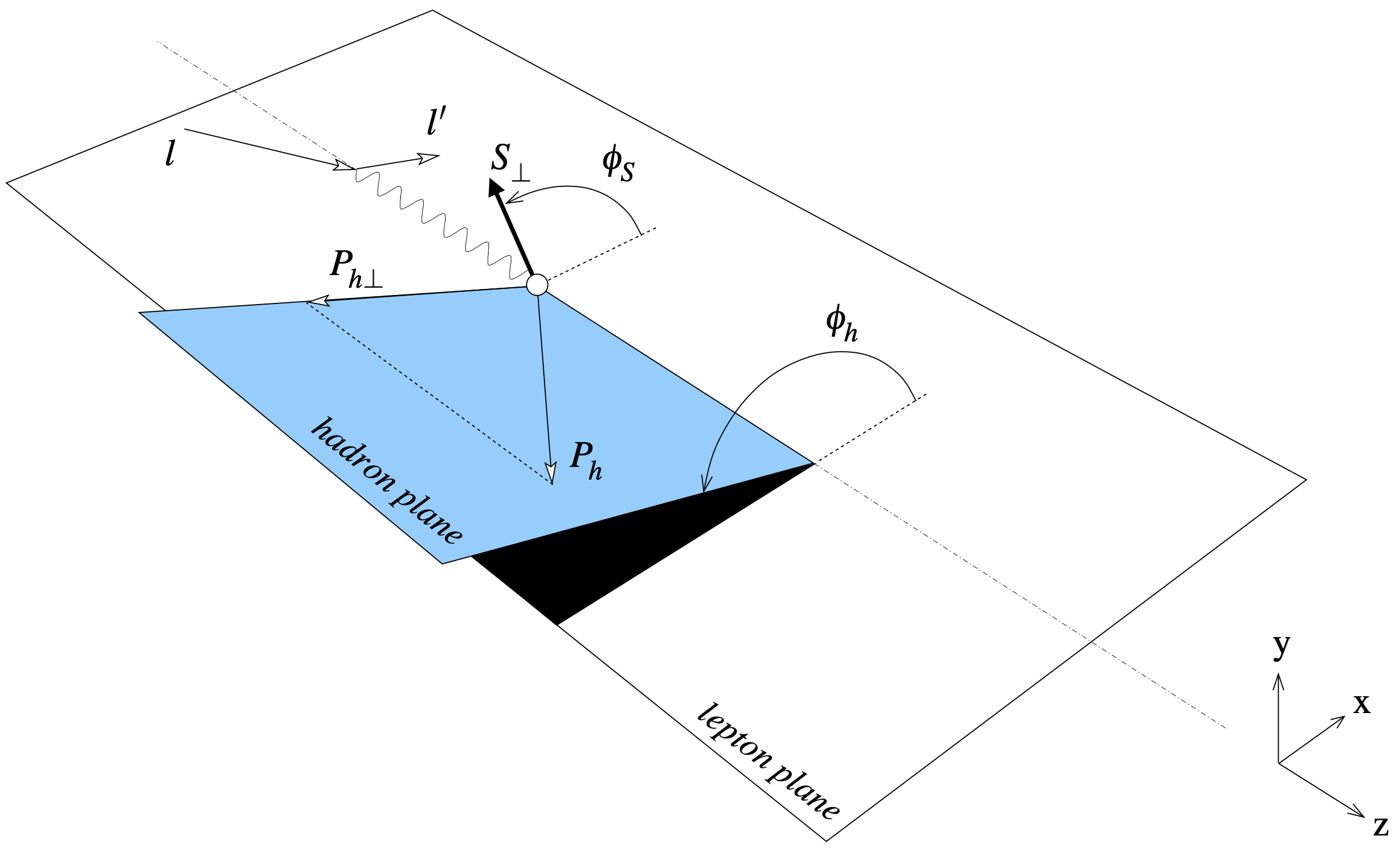}
    \caption{SIDIS scattering and azimuthal angles in the target frame~\cite{A_Bacchetta_2007}.}
    \label{fig:SIDIS_drawing}
\end{figure}

We follow the Trento conventions~\cite{PhysRevD.70.2004_bacchetta} to express the azimuthal angle $(\phi_h)$ of the outgoing hadron as shown in Fig.~\ref{fig:SIDIS_drawing} 
\begin{align}
    \sin{\phi_h}=& -\frac{l_u {(P_h)_\nu} \epsilon^{\mu\nu}_\perp}{\sqrt{l^2_\perp P^2_{h\perp}}} \qquad
    \cos{\phi_h}= -\frac{l_u {(P_h)_\nu}g^{\mu\nu}_\perp}{\sqrt{l^2_\perp P^2_{h\perp}}} 
\end{align}
\noindent where $l^\mu_\perp = g^{\mu\nu}_\perp l_\nu$ and $P^\mu_{h\perp} = g^{\mu\nu}_\perp P_{h\nu}$ are the transverse components of $\mathbf{l}$ and $\mathbf{P_h}$ with respect to the photon momentum.

Similarly $(\phi_S)$ is defined in terms of covariant spin vector \textbf{S} of the target as
\begin{align}
    \sin{\phi_S}=& -\frac{l_u S_{\nu}\epsilon^{\mu\nu}_\perp}{\sqrt{l^2_\perp S^2_{\perp}}} \qquad
    \cos{\phi_S}= -\frac{l_u S_{\nu}g^{\mu\nu}_\perp}{\sqrt{l^2_\perp S^2_{\perp}}} 
\end{align}

Considering longitudinal polarization only, we could express the differential cross-section for this Semi-Inclusive Deep Inelastic Scattering (SIDIS) including the tensor polarization terms as in Ref.~\cite{bacchetta:2023Contact}:

\begin{align}
    \frac{d\sigma}{dx~dy~d\psi~dz~d\phi_h~dP^2_{h\perp}} =& \frac{y^2\alpha^2}{2(1-\epsilon)xyQ^2} \left(1+\frac{\gamma^2}{2x}\right) \Bigg[F_{UU,T}+\epsilon F_{UU,L}+\sqrt{2\epsilon(1+\epsilon)}\cos{\phi_h} ~F_{UU}^{\cos\phi_h}\nonumber\\
    &+\epsilon \cos(2\phi_h)~F_{UU}^{\cos{(2\phi_h)}}+\lambda_e\sqrt{2\epsilon(1-\epsilon)}\sin{\phi_h}~F_{LU}^{\sin{\phi_h}}\nonumber\\
    -----&--------------------------\nonumber\\
    \text{Vector polarization:}\qquad &+S_{\parallel}\bigg\{\sqrt{2\epsilon(1+\epsilon)}\sin{\phi_h}~F_{UL}^{\sin{\phi_h}}+\epsilon\sin{(2\phi_h)}~F_{UL}^{\sin{2\phi_h}}\bigg\}\nonumber\\
    &+S_\parallel\lambda_e\bigg\{\sqrt{1-\epsilon^2}~F_{LL}+\sqrt{2\epsilon(1-\epsilon}\cos{\phi_h}~F_{LL}^{\cos{\phi_h}}\bigg\}\nonumber\\
    -----&--------------------------\nonumber\\
    \text{Tensor Polarization:}\qquad &+T_{\parallel\parallel}\bigg\{F_{U(LL),T}+\epsilon~F_{U(LL),L}+\sqrt{2\epsilon(1+\epsilon)}\cos{\phi_h}~F_{U(LL)}^{\cos{\phi_h}}\nonumber\\
    &+\epsilon\cos{(2\phi_h)}~F_{U(LL)}^{\cos{2\phi_h}}+\lambda_e\sqrt{2\epsilon(1-\epsilon)}\sin{\phi_h}~F_{L(LL)}^{\sin{\phi_h}}
    \bigg\}
    \Bigg]
    \label{eq:cross-section}
\end{align}

\noindent where $\lambda_e$ is the helicity of electron beam, $S_\parallel$ is the target vector polarization parallel to the virtual photon momentum, and $T_{\parallel \parallel}$ is the tensor polarization. 

The total cross section of a vector and tensor-polarized deuteron target is composed of the unpolarized, vector and the tensor contributions. In Eq~\ref{eq:cross-section}, the first two lines are the unpolarized contribution and the following two lines are contribution from vector polarization, weighted by the amount of vector polarization $S_\parallel$. Lastly, the tensor contribution is weighted by the size of tensor polarization, $T_{\parallel \parallel}$ in the last two lines.  The kinematic factor  $\epsilon$ is the ratio of longitudinal to transverse photon flux~\cite{A_Bacchetta_2007} written as

\begin{align}
    \epsilon =&~\frac{1-y-\frac{1}{4}\gamma^2 y^2}{1-y+\frac{1}{2}y^2+\frac{1}{4}\gamma^2y^2}
\end{align}

Contributions from the tensor polarization in the cross-section are expressed in terms of tensor polarized structure functions ($F$) which can be expressed in terms of the TMD distribution functions ($f,h,e~\text{and}~g$) and fragmentation correlation functions ($D,H,E~\text{and}~G$)~\cite{bacchetta:2023Contact,PhysRevD.103.2021_Kumano,proceeding_Kumano_2022} as following:

\begin{subequations} \label{eq:TMD_observable}
    {\footnotesize
    \begin{align}
    {F_{U(LL),T}} &= C[f_{1LL}D_1] \label{eq:TMD_observable1}\\
    {F_{U(LL),L}} &=~ 0 \label{eq:TMD_observable2}\\
    {F_{U(LL)}^{\cos{\phi_h}}} &=~ \frac{2M}{Q}C\Bigg[-\frac{\hat{\mathbf{h}}\cdot \mathbf{k_T}}{M_h}\bigg(xh_{LL}H_1^\perp+\frac{M_h}{M}f_{1LL}\frac{\Tilde{D}^\perp}{z}\bigg)-\frac{\hat{\mathbf{h}}\cdot \mathbf{p_T}}{M}\bigg(xf_{LL}^\perp D_1 +\frac{M_h}{M}h_{1LL}^\perp\frac{\Tilde{H}}{z}\bigg)\bigg] \label{eq:TMD_observable3}\\
    {F_{U(LL)}^{\cos{2\phi_h}}} &=~ C\Bigg[-\frac{2(\hat{\mathbf{h}}\cdot \mathbf{k_T})(\hat{\mathbf{h}}\cdot \mathbf{p_T})-\mathbf{k_T}\cdot \mathbf{p_T}}{MM_h}h_{1LL}^{\perp} H_1^{\perp} \bigg] 
    \label{eq:TMD_observable4} \\
    {F_{L(LL)}^{\sin{\phi_h}}} &=~ \frac{2M}{Q}C\Bigg[-\frac{\hat{\mathbf{h}}\cdot \mathbf{k_T}}{M_h}\bigg(xe_{LL}H_1^\perp+\frac{M_h}{M}f_{1LL}\frac{\Tilde{G}^\perp}{z}\bigg)+\frac{\hat{\mathbf{h}}\cdot \mathbf{p_T}}{M}\bigg(xg_{LL}^\perp D_1 +\frac{M_h}{M}h_{1LL}^\perp\frac{\Tilde{E}}{z}\bigg)\bigg] \label{eq:TMD_observable5}
    \end{align}
    }
\end{subequations}
where $\hat{\mathbf{h}}=\frac{\mathbf{P}_{h\perp}}{|{\mathbf{P}_{h\perp}}|}$ and with an arbitrary funtion $w(\mathbf{p}_T,\mathbf{k}_T$), we have
\begin{align}
    C =&~x\sum_{a}^{} e_a^2 \int d^2\mathbf{p}_T~d^2\mathbf{k}_T~\delta^2(\mathbf{p}_T - \mathbf{k}_T - \mathbf{P}_{h\perp}/z)~w(\mathbf{p}_T,\mathbf{k}_T)~f^a(x,p_T^2)~D^a_h(z,k_T^2)
\end{align}
with sum over quarks and anti-quarks~\cite{A_Bacchetta_2007}.

In this exploratory analysis, we will attempt to access the ${F_{U(LL),T}}$ and $F_{U(LL)}^{\cos{2\phi_h}}$ structure functions, which provide access to the $f_{1LL}$ and $h_{1LL}^\perp$ TMDs in a clean way without interference from other distribution functions. By carefully choosing the hadron angle $\phi_h$, using world unpolarized data, and testing CLAS12 data with equal tensor polarization and opposite vector polarization, we can isolate the individual terms of the tensor polarization part of Eq~\ref{eq:cross-section} to extract the structure functions of interest. These observables will allow us to study QCD in a more refined way and also provide novel information on deuteron about the interplay between QCD and nuclear structure.

\subsubsection*{Inclusive DIS tensor structure functions}
Among the various tensor structure functions expressed in Eq.~\ref{eq:TMD_observable}, $F_{U(LL),T}$ is related with the inclusive DIS structure function $b_1$~\cite{A_Bacchetta_2007,PhysRevD.103.2021_Kumano}. In partonic model, the inclusive tensor structure functions $b_1$ and $b_2$ can be expressed as: 
%
\begin{eqnarray}
b_1(x) &=& \frac{1}{2} \sum_q e_q^2\Big( 2 q^0_{\uparrow}(x) - q^1_{\uparrow}(x) - q^1_{\downarrow}(x) \Big) \\
b_2(x) &=& 2 x b_1(x)
\label{TSF-parton}
\end{eqnarray}
where $q^m_{\uparrow}$ ($q^m_{\downarrow}$)  represents the probability to find a quark with momentum fraction $x$ and spin up (down) in a hadron which is in helicity state $m$.
%
The tensor structure function $b_1$ depends only on the spin-averaged parton distributions since, by parity, $q^m_{\uparrow}= q^{-m}_{\downarrow}$. 
\begin{eqnarray*}
q^1(x) &=& q^1_{\uparrow}(x) + q^1_{\downarrow}(x) = q^1_{\uparrow}(x) + q^{-1}_{\uparrow}(x)\\ 
q^0(x) &=& q^0_{\uparrow}(x) + q^0_{\downarrow}(x) 
= 2 q^0_{\uparrow}(x)
\end{eqnarray*}
so it can be expressed as:
\begin{eqnarray}
b_1(x) = \frac{1}2\sum_q e_q^2 [q^0(x) - q^1(x)]
\end{eqnarray}

Explicitly, $b_1$ measures the difference
in partonic constituency in an $|m|$=1 target and an $m$=0 target. 
 From this we see that while $b_1$ is defined in terms of quark distributions, it  depends also on the spin state of the nucleus as a whole.

 The integral of b1 is described by the sum role of Close and Kumano~\cite{PhysRevD.42.1990_CloseKumano}:

\begin{eqnarray}
\int_0^1 dx~b_1(x)  & = &  - \frac{5}{12 M^2} \lim_{t \rightarrow 0}~t~F_Q(t)
                           + \frac{1}{9} \Big(\delta Q + \delta \bar{Q}\Big)_s \nonumber \\
                    & = & \frac{1}{9} \Big(\delta Q + \delta \bar{Q}\Big)_s  = 0 
\label{cksum}
\end{eqnarray}
where $F_Q(t)$ is the electric quadrupole form factor of a spin-1 hadron at the momentum squared $t$. 
The Close Kumano (CK) sum rule is satisfied in the case of an unpolarized sea. The authors note
that in nucleon-only models, the integral of $b_1$ is not sensitive to the
tensor-polarization of the sea, and consequently the sum rule is always true, even when the
deuteron is in a $D$-state.

\subsection{Previous Measurements}

The deuteron is the simplest nucleus with spin-1, and both theoretical and experimental efforts are ongoing to understand it in more detail. In addition to the study of unpolarized and polarized deuteron structure functions, the tensor structure functions of the deuteron have interesting features potentially relevant to exotic effects in nuclei~\cite{PhysRevC.44.1991}.

\begin{figure}[htb]
    \centering
    \includegraphics[height=11cm,width=10cm]{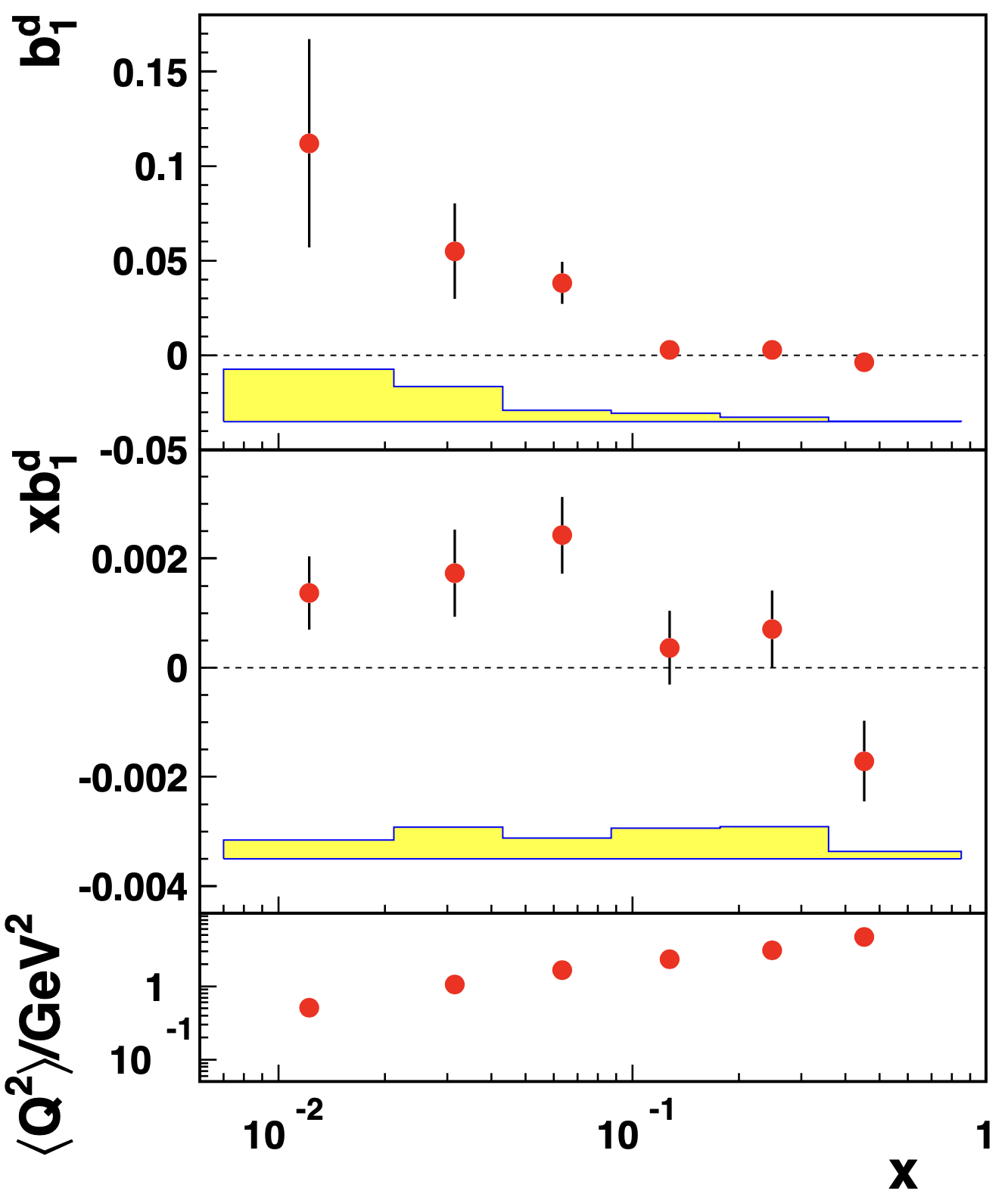}
    \caption{HERMES results on the tensor structure function $b_1$ with experimental uncertainties and showing a crossover~\cite{PhysRevLett.95.2005_Hermes}.}
    \label{fig:hermes_b1}
\end{figure}

\begin{figure}[htb]
    \centering
    \includegraphics[height=8cm,width=10cm]{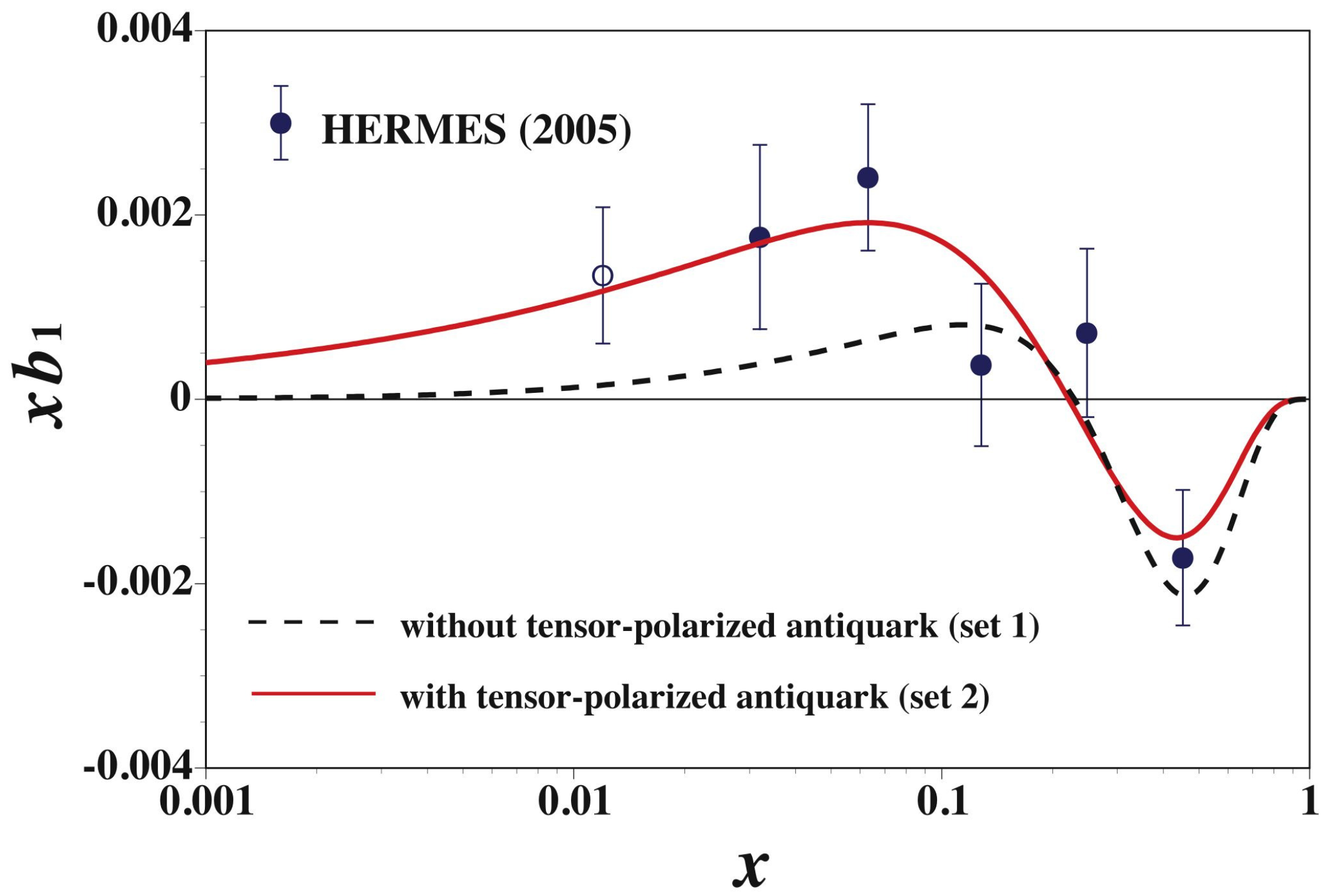}
    \caption{HERMES data fitted with theoretical model in Ref.~\cite{PhysRevD.82.2010_Kumano} with and without tensor polarized antiquark distribution.}
    \label{fig:kumano_fit}
\end{figure}

\begin{figure}[htb]
    \centering
    \includegraphics[height=8cm,width=10cm]{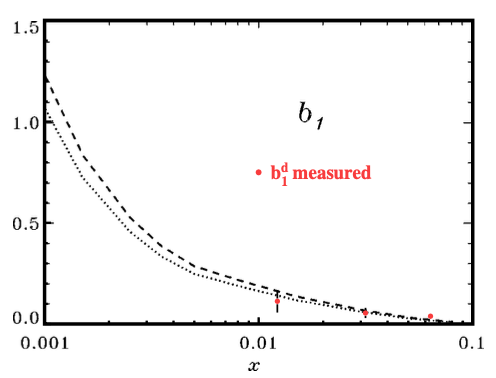}
    \caption{A dramatic increase in the value of $b_1$ at low x is predicted from shadowing effects.  The red data points are from HERMES~\cite{PhysRevLett.95.2005_Hermes}. }
    \label{fig:lowx}
\end{figure}

The tensor structure function of the deuteron ($b_1$) was first measured experimentally by the HERMES Collaboration, with a positron beam impinging on a tensor polarized deuteron gas target. All conventional nuclear physics models, which includes effects beyond a free proton and neutron, predict $b_1$ to be very small ($<10^{-3}$) at moderate $x\gtrsim 0.2$.
The data from HERMES~\cite{PhysRevLett.95.2005_Hermes} revealed a significantly larger negative value of $b_1$ in this region, albeit with relatively large uncertainty as shown in Fig.~\ref{fig:hermes_b1}.
The non-zero value of $b_1$ indicated by this data also points towards the tensor polarized anti-quark distribution at low $x$ as shown in Fig.~\ref{fig:kumano_fit} and explained in Ref.~\cite{PhysRevD.82.2010_Kumano}. Furthermore, $b_1$ is expected to be small at moderate x but quite large at low x due to shadowing as shown in Fig.~\ref{fig:lowx}.

A new experiment is approved to measure the tensor structure function ($b_1$) in the Hall C of Jefferson lab via the tensor asymmetry, $A_{zz}$, utilizing existing HMS and SHMS spectrometers, which will measure $b_1$ more precisely and map out the $x$-region where a zero-crossing is expected~\cite{proposal_b1_jlab}. There has been good progress on the tensor polarized target development and the tentative accelerator schedule projects this experiment to run in the next few years, in parallel with the M\"oller experiment. There are  experimental programs anticipated  at various  facilities~\cite{PhysRevD.103.2021_Kumano,proceeding_Kumano_2022} that may be able to access $b_1, b_2, b_3$, and $b_4$ in the future, but at present JLab is the only facility with the capability and plans to do so. 
\section{Data}
\label{section:data}
We will analyze the data from the CLAS12 Run Group C (RG-C), which took data on longitudinally polarized ammonia and deuterated ammonia targets with polarized electron beams during 2022 and 2023. RG-C comprises five PAC-approved experiments to study spin structure functions, Generalized Parton Distributions (GPDs) and TMDs with a longitudinally polarized target, as well as three other run group experiments~\cite{HallB_exp,RGC_exp}.

\begin{figure}[htb]
    \centering
    \includegraphics[height=12cm,width=16cm]{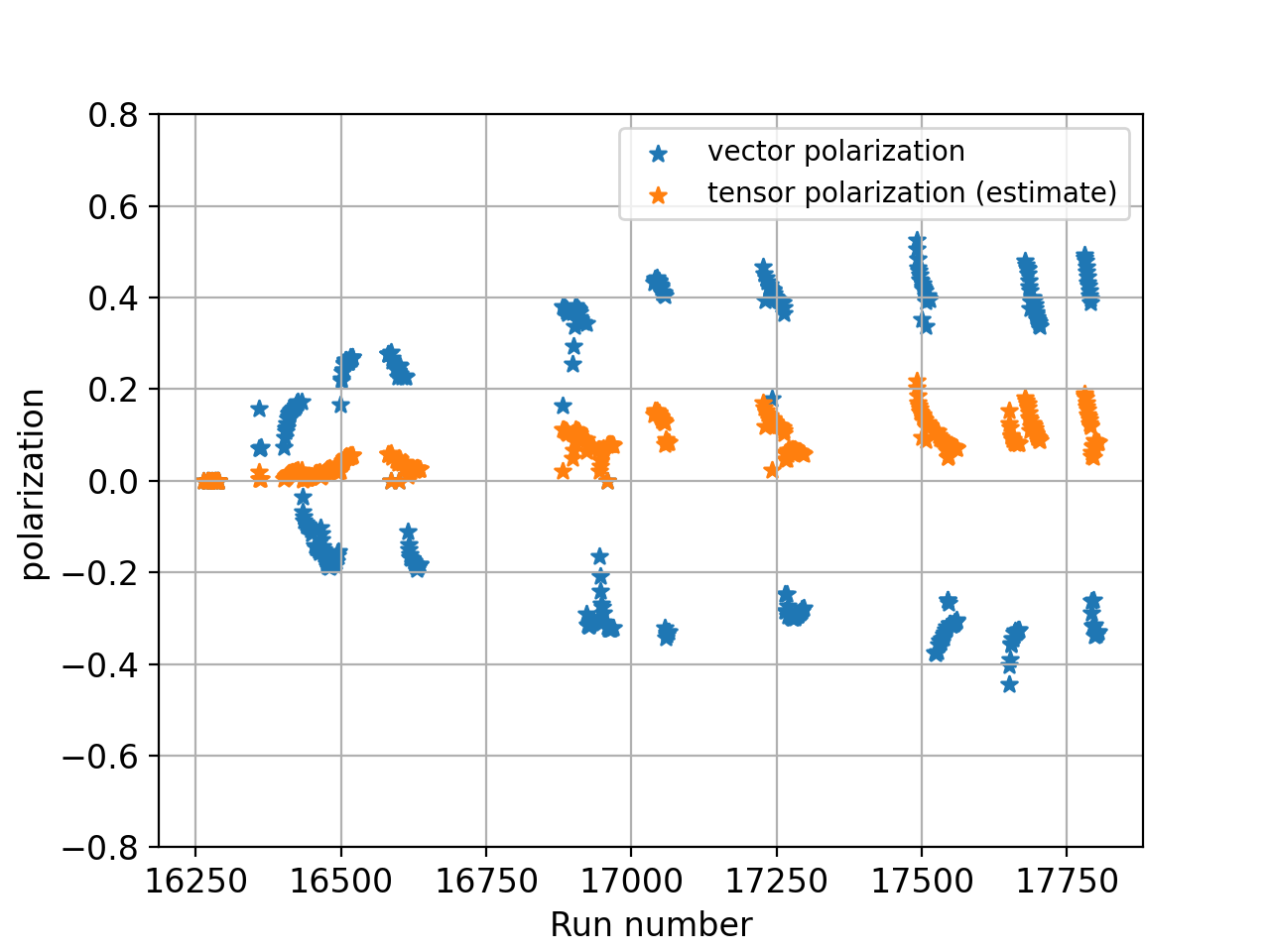}
    \caption{Vector polarization ($+S_{\parallel}$ and $-S_{\parallel}$) (blue) of  the ND$_3$ target during the RG-C experiment with new calibration constants for the NMR measurements. Estimated tensor polarization ($T_{\parallel \parallel}$) (orange) of ND$_3$ target based on the vector polarization numbers~\cite{CLAS12_target}. See text for more information.}
    \label{fig:nd3_pol_vecten} 
\end{figure}

During the RG-C run period, the target polarization of the deuterated ammonia was measured via Nuclear Magnetic Resonance (NMR). The average longitudinal polarization of the deuterated ammonia (ND$_3$) target was $31.0\%$ in the positive $S_{\parallel}$ direction and $25.5\%$ in the negative $S_{\parallel}$ direction after evaluating the calibration constant from the offline analysis~\cite{CLAS12_target,RGC_data}.
This polarization is shown by run in Fig.~\ref{fig:nd3_pol_vecten}.

This vector polarization ($S_\parallel$) is related with the tensor polarization ($T_{\parallel\parallel}$) of deuterium in thermal equilibrium with the solid lattice by 
\begin{align}
    T_{\parallel\parallel} =& 2- \sqrt{4-3S_\parallel^2}
\end{align}
neglecting the small quadrupole interaction. This relation implies that a sizable amount of tensor polarization is available in the RG-C data and currently is not being used for any analysis. An initial estimation of this tensor polarization based on the available vector polarization numbers is shown in Fig.~\ref{fig:nd3_pol_vecten}.

\begin{figure}[htb]
    \centering
    \includegraphics[height=8.5cm,width=8.6cm]{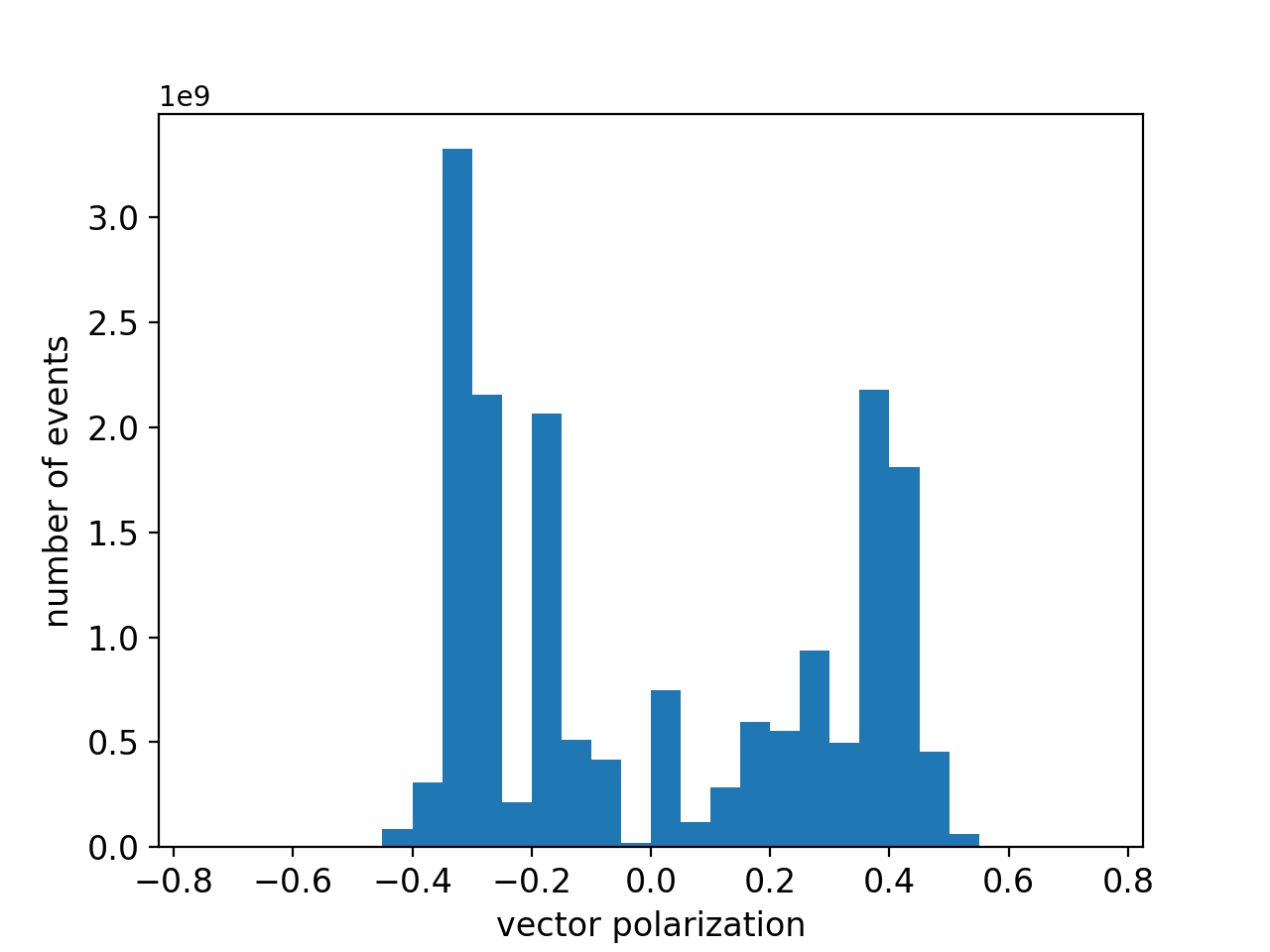} \hspace{-1cm}
    \includegraphics[height=8.5cm,width=8.6cm]{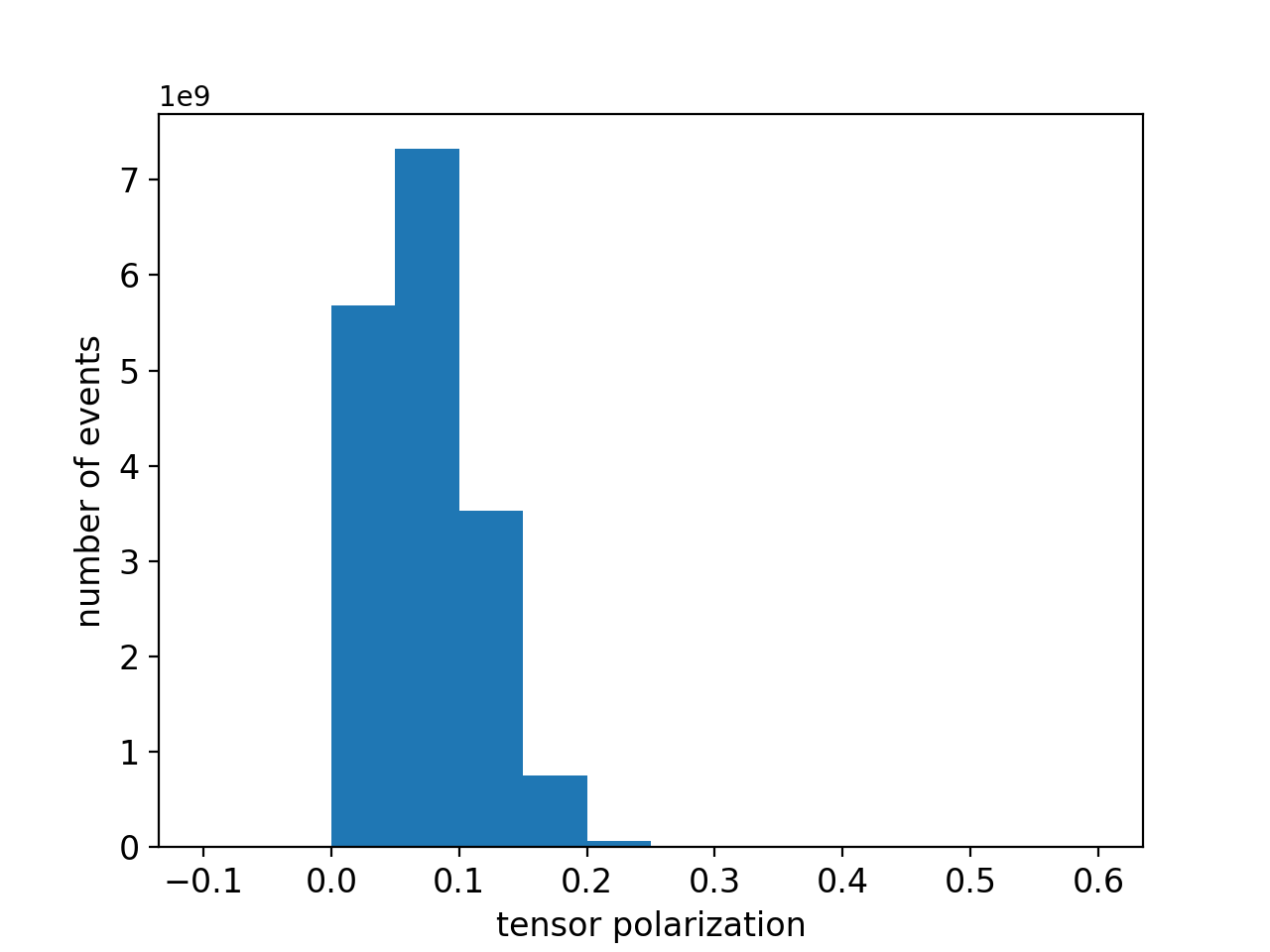}
    \caption{Number of vector polarization ($\pm S_{\parallel}$) events on the deuterated ammonia target during the RG-C experiment on the left. Estimated number of events with tensor polarization ($T_{\parallel \parallel}$) on the right~\cite{CLAS12_target,RGC_data}.}
    \label{fig:nd3_TotalNevents}
\end{figure}

As shown in Fig.~\ref{fig:nd3_pol_vecten}, RG-C took data with both $+S_{\parallel}$ and $-S_{\parallel}$ polarity, but without any runs on unpolarized ND$_3$. There are about 900 million events on each polarity, as shown in Fig.~\ref{fig:nd3_TotalNevents}, which will be used for our exploratory study on tensor structure functions.

In addition to production data on polarized NH$_3$ and ND$_3$ targets during the RG-C run period, data on auxiliary targets, including empty, C$_{12}$, CH$_2$, and CD$_2$ targets, were taken to determine the dilution factor. The dilution factor corrects contributions to the data from electron scattering off other materials in the target besides deuterons. Currently, it is estimated that the dilution factor is of the order of $\sim$0.27~\cite{RGC_dilution}.

\section{Proposed analysis}
\label{section:analysis} 

We will use the CLAS12 RG-C polarized ND$_3$ data to study the tensor structure functions of spin-1 hadrons via SIDIS and inclusive analyses. For SIDIS analysis, we will use events with pions in the final states; $eD \rightarrow e'\pi^{+}X$ and $eD \rightarrow e'\pi^{-}X$.
The outcome of the this analysis will be the observables $F_{U(LL),T}$ and $F_{U(LL)}^{\cos{2\phi_h}}$ expressed in Eq.~\ref{eq:TMD_observable}. Despite the low tensor polarization in RG-C, we expect this analysis will be the first ever extraction of these quantities, and will provide both novel results in its own right, and strong motivation and guidance for future dedicated measurements of these quantities.

\begin{figure}[htbp]
    \centering
    \includegraphics[height=8cm,width=14cm]{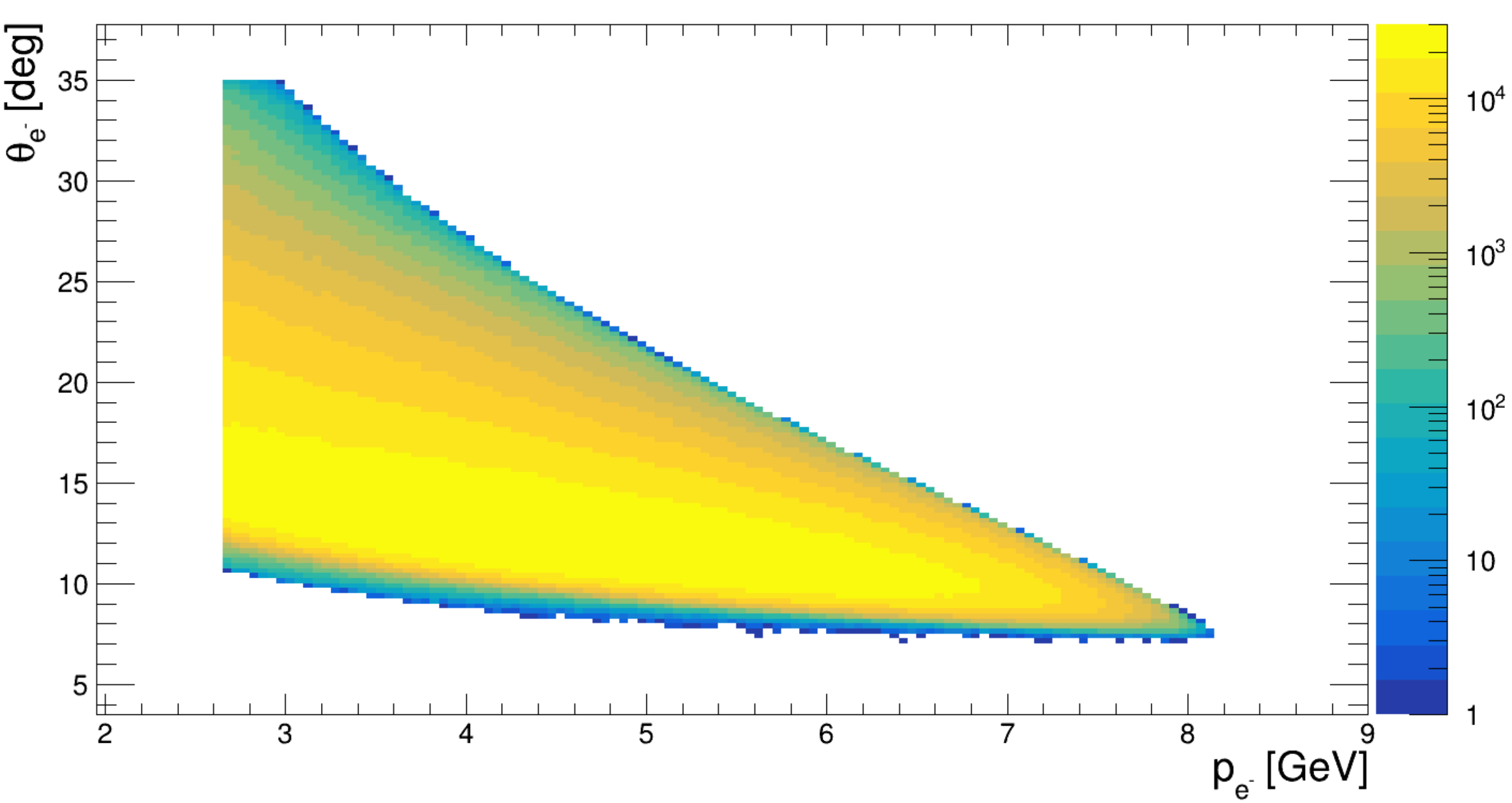}
    \includegraphics[height=8cm,width=14cm]{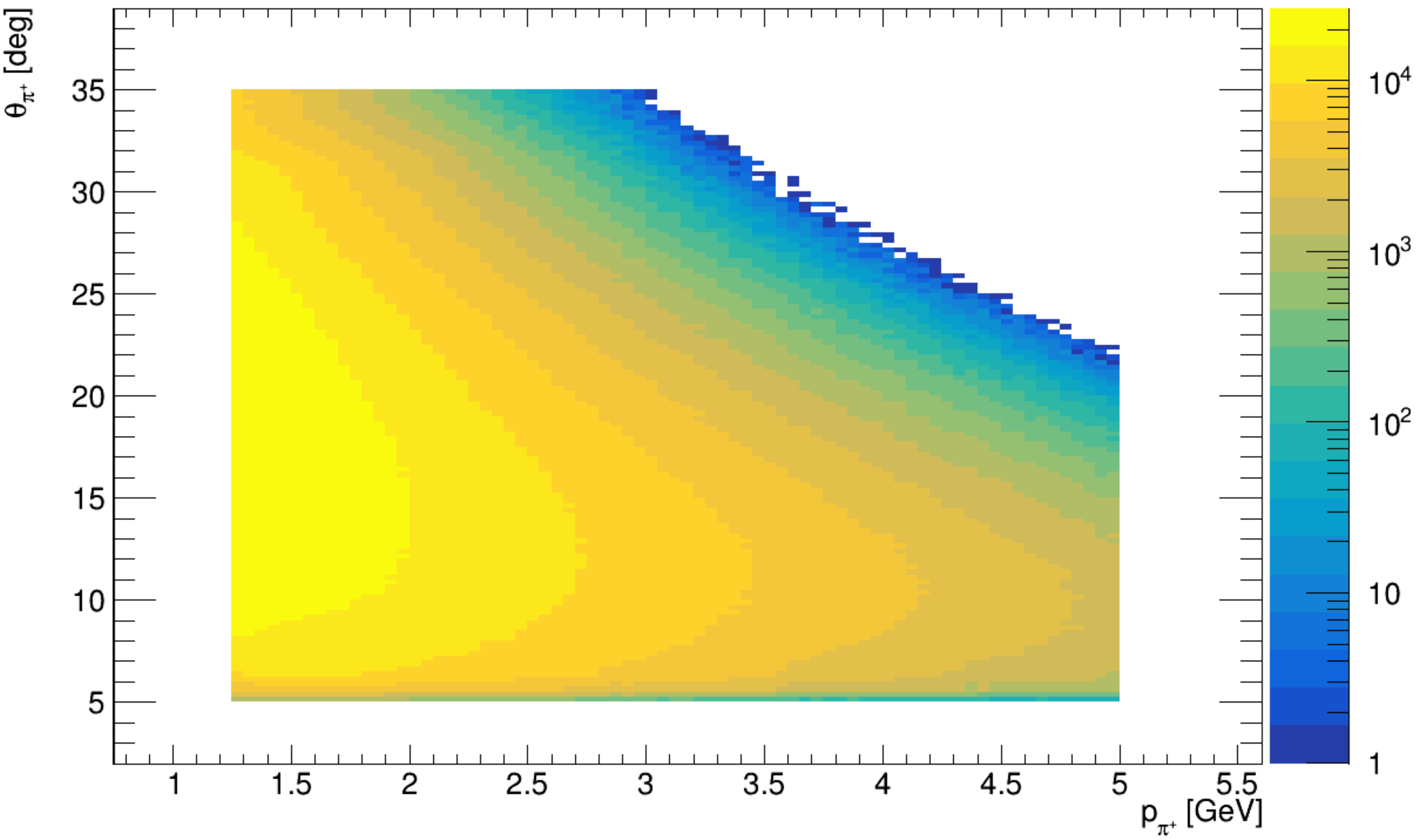}
    \caption{Polar angular acceptance of the CLAS12 forward detector for e$^-$ (top) and $\pi^+$ (bottom) with inbending torus field obtained from the SIDIS Beam Spin Assymetry (BSA) analysis of CLAS12 RG-A data~\cite{RGA_Analysis_note_Diehl}.}
    \label{fig:theta_coverage_RGA_SIDIS}
\end{figure}

The selected events will have one scattered electron and pion in the final state, both detected in the forward detector of the CLAS12 spectrometer. This forward detector covers the polar angle from $5^\circ$ to $35^\circ$ as illustrated in Fig.~\ref{fig:theta_coverage_RGA_SIDIS} with the CLAS12 Run Group A (RG-A) experimental data.

\begin{figure}[htbp]
    \centering
    \includegraphics[height=13.5cm,width=16cm]{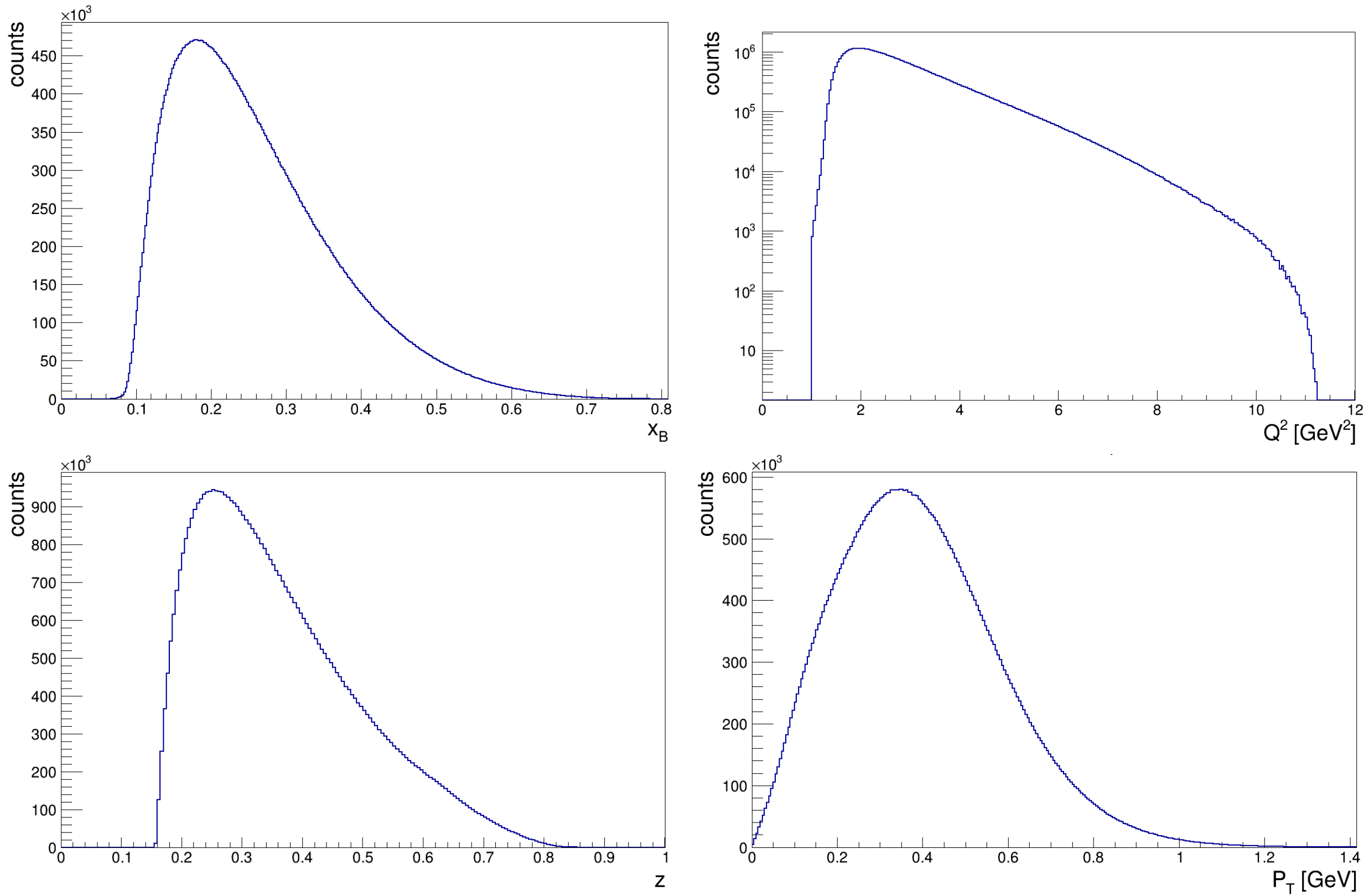}
    \caption{Distribution of SIDIS kinematic variables ($x_B, Q^2, z~\text{and}~P_T$) achievable in CLAS12 for $ep \rightarrow e\pi^+X$ analysis~\cite{RGA_Analysis_note_Diehl}.}
    \label{fig:kinematics_RGA_SIDIS_1}
\end{figure}

\begin{figure}[htbp]
    \centering
    \includegraphics[height=8cm,width=12.5cm]{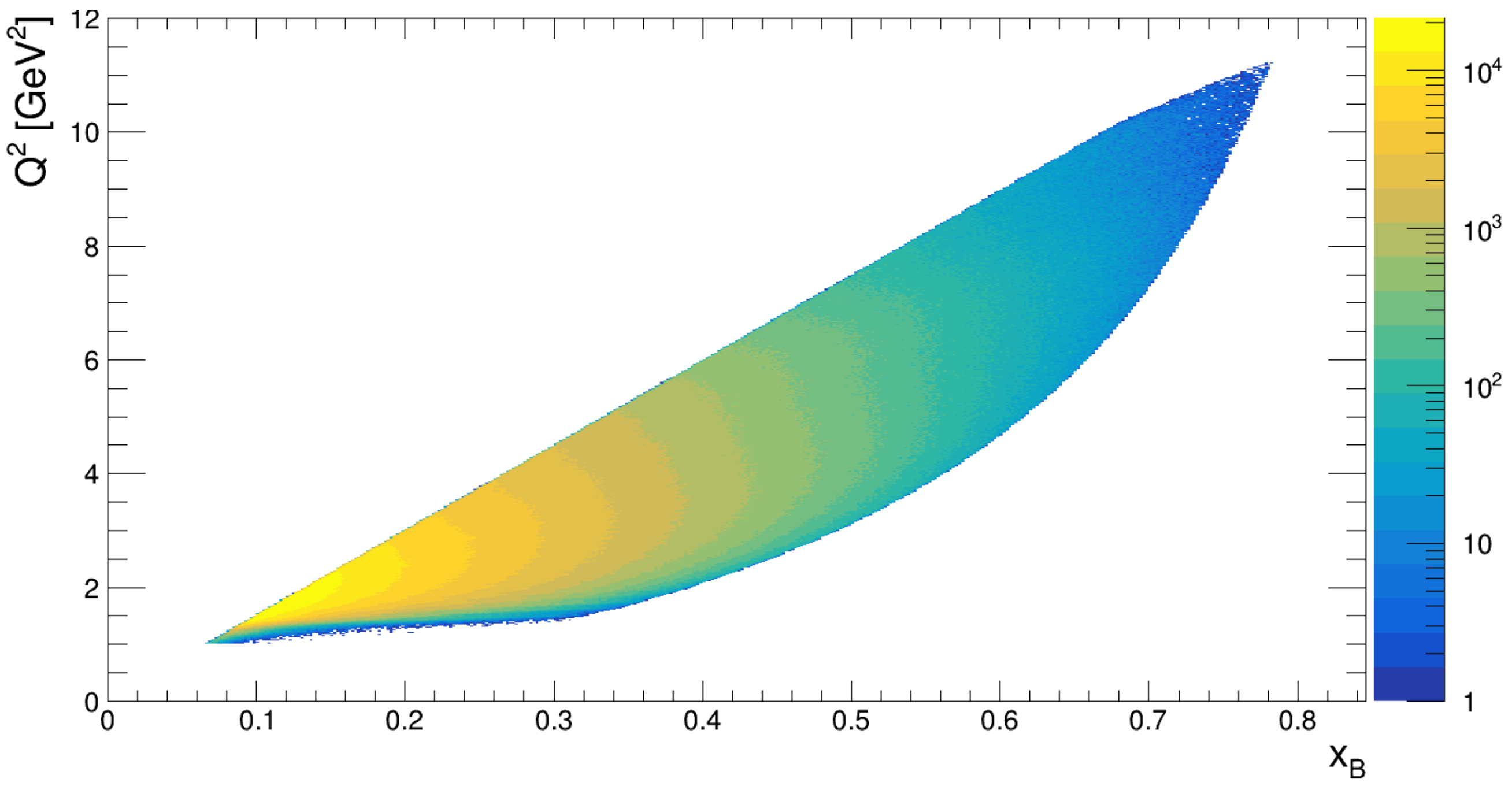}
    \caption{Distribution of $Q^2~\text{vs}~x_B$ for SIDIS analysis of $ep \rightarrow e\pi^+X$ in CLAS12~\cite{RGA_Analysis_note_Diehl}.}
    \label{fig:kinematics_RGA_SIDIS_2}
\end{figure}

\begin{figure}[htbp]
    \centering
    \includegraphics[height=6.5cm,width=10cm]{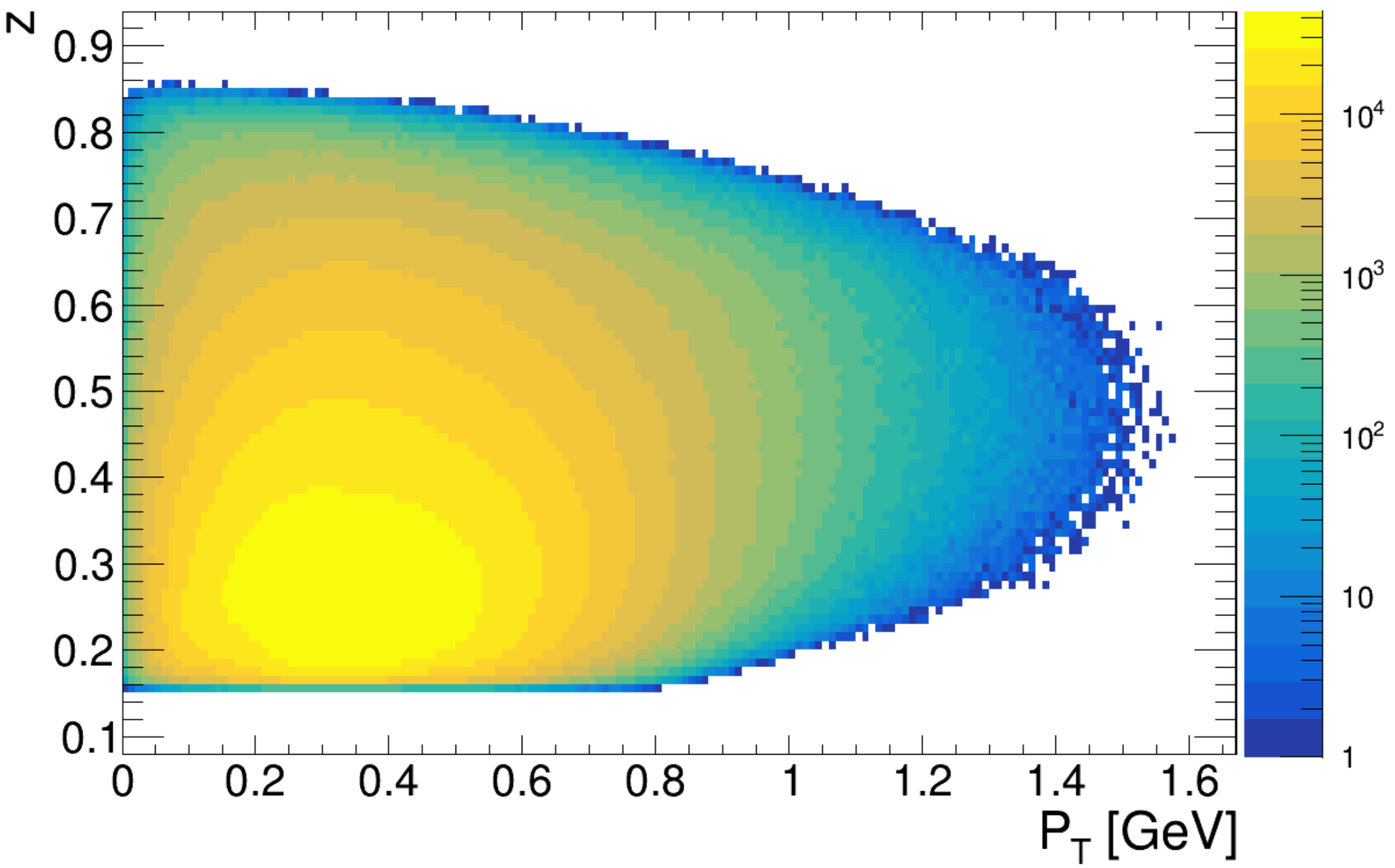}
    \caption{Distribution of $z~\text{vs}~P_T$ for SIDIS analysis of $ep \rightarrow e\pi^+X$ in CLAS12~\cite{RGA_Analysis_note_Diehl}.}
    \label{fig:kinematics_RGA_SIDIS_3}
\end{figure}

With the electrons and pions detected in the final state, the kinematic coverage of various variables that could be achieved in the CLAS12 for the SIDIS analysis is shown in Fig~\ref{fig:kinematics_RGA_SIDIS_1}, \ref{fig:kinematics_RGA_SIDIS_2} and \ref{fig:kinematics_RGA_SIDIS_3}, which are available from the SIDIS analysis of RG-A data~\cite{RGA_Analysis_note_Diehl}. These plots show that CLAS12 has a large coverage on $Q^2$, $x_B$, $z$ and $P_T$ with 11.6 GeV beam. In this analysis, we will use the data obtained with a longitudinally polarized ND$_3$ target in RG-C and we expect the CLAS12 kinematic coverage will be similar. There were no major changes in the spectrometer in these two experiments, even though RG-A used liquid hydrogen target, to get a dramatically different kinematic coverage.

Additionally, we will also use these data for the inclusive analysis to extract the collinear function $b_1$. The measurement we will obtain from this dataset for this parameter will be the second measurement ever conducted.

\subsection{Analysis plan}
The SIDIS cross-section will have unpolarized, vector polarized, and tensor polarized contributions as detailed in equation~\ref{eq:cross-section}. Therefore, we have the following plan to extract the $F_{U(LL),T}$ and $F_{U(LL)}^{\cos{2\phi_h}}$ contributions to the cross sections:

\begin{itemize}

\item \textbf{Suppression of vector contribution ($S_{\parallel}$})  

Contribution of the longitudinal vector contribution to the total cross section will be suppressed by combining data-sets with positive and negative longitudinal target polarization with proper weighting.

\item \textbf{Suppression of electron helicity  contribution ($\lambda _e$})  

The electron beam in Jefferson Lab has a polarized component attached to it, and RG-C has collected beam polarization measurements several times over the run period. To remove the contribution of $\lambda_e$ in the tensor terms, we will sum over events with both positive and negative beam polarization with proper weighting.

\item \textbf{Dilution factor}

Currently, the dilution factor for RG-C is estimated around $\sim$0.27~\cite{RGC_dilution}, but a detailed study will be performed to remove the contribution to the data from other materials along beamline other than deuterons.

\item \textbf{Extraction of tensor part}       

RG-C did not take unpolarized ND$_3$ data, and no such data is available in CLAS12 yet. However, the leftover cross-section, after the suppression of the contribution from vector polarized target and beam helicity, will be plotted as a function of tensor polarization dividing it into various bins as shown in Appendix~\ref{appendix:A}. Slope of the linear fit of these data points would be the tensor polarized cross-section with suppressed unpolarized contribution. 


After the steps above, we will have contributions only from tensor parts as in Eq.~\ref{eq:cross-section_tensor}: 

\begin{align}
    \frac{d\sigma}{dx~dy~d\psi~dz~d\phi_h~dP^2_{h\perp}} =&\frac{y^2\alpha^2}{2(1-\epsilon)xyQ^2} \left(1+\frac{\gamma^2}{2x}\right)T _{\parallel\parallel}\nonumber\\
    &\bigg\{F_{U(LL),T} + \epsilon\cos{(2\phi_h)}~F_{U(LL)}^{\cos{2\phi_h}}++\sqrt{2\epsilon(1+\epsilon)}\cos{\phi_h}~F_{U(LL)}^{\cos{\phi_h}}
    \bigg\}
    \label{eq:cross-section_tensor}
\end{align}

The first term $F_{U(LL)T}$ will be extracted by integrating over $\phi_h$ after following all the above steps. In addition, the extraction of $F_{U(LL)}^{\cos{2\phi_h}}$ will be performed with the angular modulation by $\cos{(2\phi)}$.

\end{itemize}

\begin{figure}[htbp]
    \centering
    \includegraphics[height=7.5cm,width=12cm]{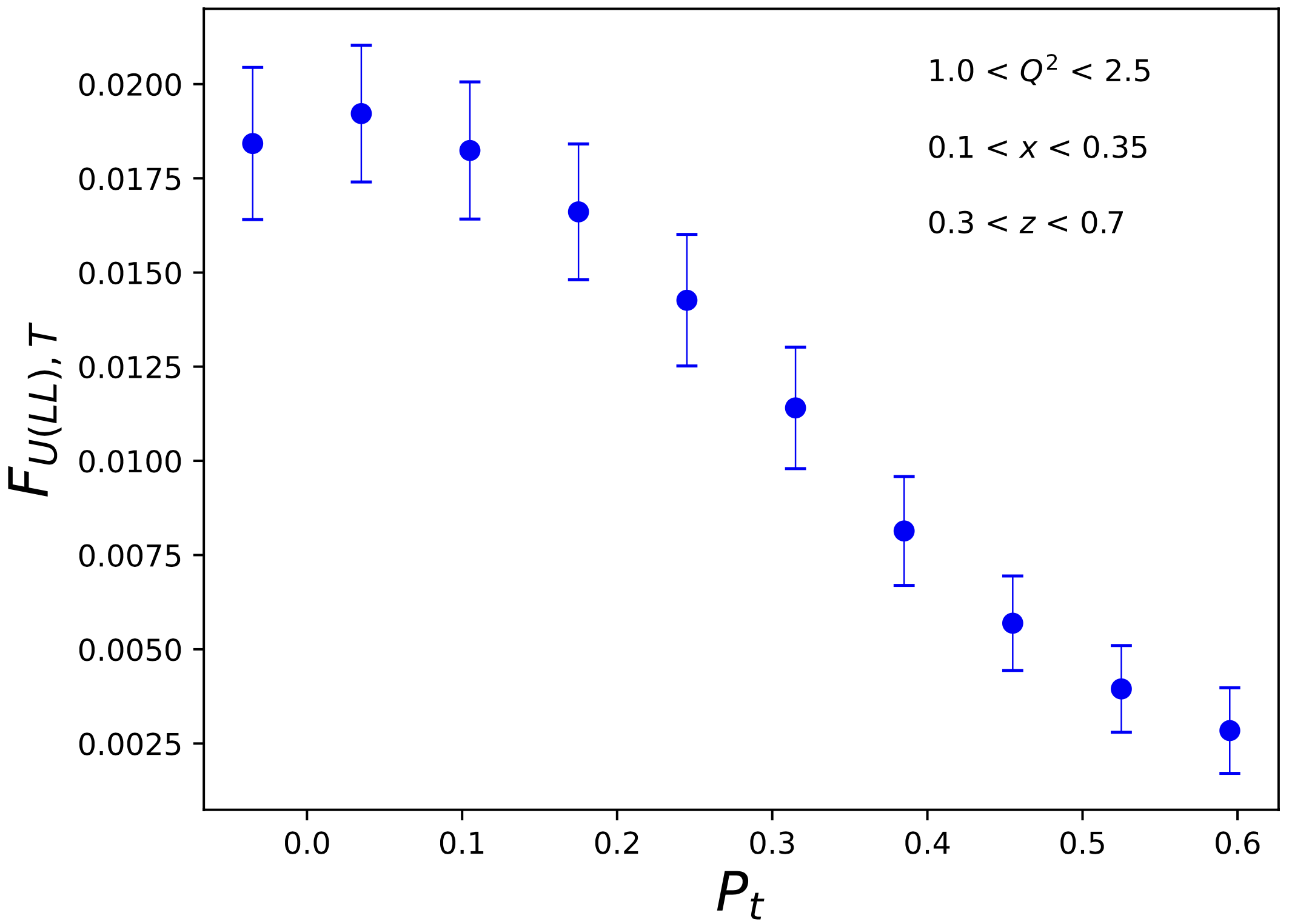}
    \caption{Projected results for the $F_{U(LL),T}$ structure function considering $\sim10\%$ of the unpolarized structure function $F_{UU,T}$ generated using LHAPDF information.}
    \label{fig:FUULT_sim}
\end{figure}

Based on the information of parton distribution function and fragmentation function from LHAPDF, unpolarized structure function ($F_{UU,T}$) is generated over the kinematic region we are interested. The tensor structure function ($F_{U(LL),T}$) is estimated to be around 10\% of the unpolarized component ($F_{UU,T}$). Figure~\ref{fig:FUULT_sim} shows the estimated result of tensor structure function within the kinematic region in this analysis.

Moreover, we will use a similar analysis process to extract the collinear function $b_1$ with the inclusive events near the zero-crossing region as shown in Fig.~\ref{fig:hermes_b1}.

\subsection{Uncertainties and Systematic Control}

To minimize uncertainty contributions, we are planning a comprehensive analysis of experimental systematics. This will include an in-depth examination of detector drifts and false asymmetries. 
Such a rigorous analysis is essential not only for optimizing the utilization of data concerning tensor observables but also for significantly enhancing the analysis efforts of the ongoing Run Group C.

Furthermore, we anticipate significant improvements through the implementation of the second round of Run Group C polarization analysis. The initial round was successfully conducted by the target groups at Jefferson Lab (JLab), the University of Virginia (UVA), along with graduate student help from Old Dominion University (ODU). The forthcoming round will incorporate advanced techniques from UVA, utilizing Deep Neural Networks (DNN) specifically designed to reduce uncertainties in tensor polarization measurements from Nuclear Magnetic Resonance (NMR) signals. These state-of-the-art techniques are expected to refine our analysis and lead to more precise results in the polarization data analysis reducing the overall uncertainties in Run Group C.

\subsection{Phase two of RG-C Polarization Data Analysis}
We implement a novel approach to enhance signal-to-noise ratios with sensitivity to the operational parameters range of polarization measurements in Nuclear Magnetic Resonance (NMR) during Run Group C  for the deuteron target. This method employs machine learning (ML) technologies to achieve reliable offline polarization extraction, significantly improving the overall figure of merit for experiments.

\begin{figure}[H]
    \centering
    \includegraphics[height=7.0cm,width=15cm]{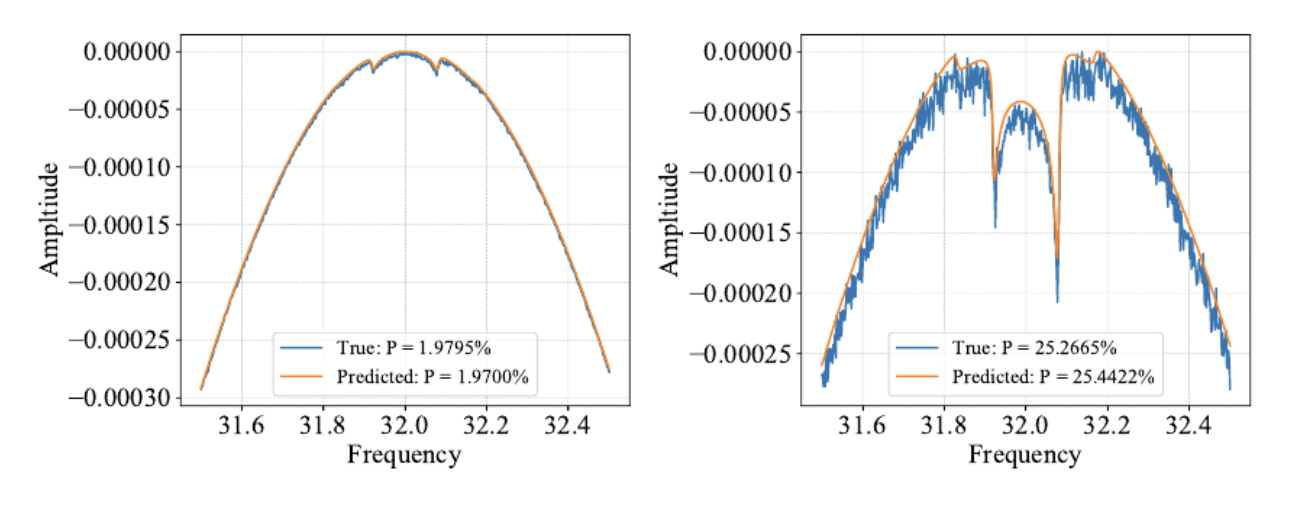}
    \caption{Example of the deuteron signal extraction using the DNN measurement method using Monte Carlo simulated target polarization data.}
    \label{DNN-det}
\end{figure}

The traditional use of Q-meters are generally limited by unknown changes in the characterization of the coil, cables, or Q-meter, or even slight deviations in the operational parameters or unrecorded RF environment variations over the course of the experiment leading to larger error in the polarization measurements. To overcome these limitations, we record all of these parameters along with environment sensor input incorporated into the neural network fitting technology used in the measurement process. Neural networks are adept at learning complex relationships between inputs and outputs, which makes them highly effective for tasks requiring high accuracy and precision. By integrating neural networks, we can now process polarization data with minimal systematic uncertainty reaching the Q-meter hardware limit of 0.8-1\% in the appropriate conditions, significantly reducing the fit error in measurements, historically around 4-6\%—to achieve more reliable results.  Figure \ref{DNN-det} shows the DNN fits to simulated polarized target data showing minimal extraction error from the new technique resulting in significant improvement to both accuracy and precision.  During the polarized target data collection some, but not all of these parameters were recorded providing an excellent opportunity to apply this approach.

There are still other limitations defined by the beam location on the target and proximity of the coil to the actual target material.  All of these factors will add additional uncertainty but with careful analysis and modeling we intend to quantify and minimize the additional error accurately.

The tensor polarization in particular is historically challenging to measure.  But the DNN approach combined with the lineshape information provides direct extraction without any reliance on the vector polarization so the error prorogation from vector measurements is completely eliminated. Through this technique, we aim to explore a new standard for polarization measurements in polarized target experiments, with the intention of facilitating more accurate and robust data analysis for this and future experiments.

\subsection{Analysis Team}

The SIDIS analysis will primarily involve J. Poudel, a postdoctoral fellow at Jefferson Lab with experience in software and CLAS12 data analysis protocols. Joining the effort is graduate student Hector Espino from the University of New Hampshire (UNH), who will be supervised and assisted by UNH postdoc D. Ruth. E. Long, D. Keller, and I. P. Fernando are tasked with polarimetry analysis, with E. Long specifically focusing on the inclusive tensor asymmetry data. K. Slifer will collaborate with graduate student Anchit Arora on the inclusive $b_1$ structure function extraction. Overseeing the entire analysis process are N. Santiesteban (UNH), D. Keller (UVa), and J. P. Chen (JLab).
\section{Summary}
\label{section:summary}

We propose the first exploratory study of spin-1 tensor structure functions, analyzing the polarized deuterated ammonia data from RG-C. The channels that we will analyze are $eD \rightarrow e'\pi^{+}X$ and $eD \rightarrow e'\pi^{-}X$ to extract the tensor structure functions. This will provide the very first measurement of the $F_{U(LL),T}$ and $F_{U(LL)}^{\cos{2\phi_h}}$ functions. In addition, we will extract the collinear $b_1$ function from the inclusive data. This analysis will generate novel results by performing this first-of-its-kind extraction, and will be crucial for future dedicated measurements to obtain a precision evaluation of these quantities. Understanding these structure functions and distribution functions is imperative if we are to fully understand the spin structure of the deuteron, and beyond it, light spin-1 nuclei in general.  These observables will provide novel information about the interplay between QCD and nuclear structure on deuteron. 

\bibliographystyle{JHEP}
\bibliography{references.bib}

\appendix
\section{Preliminary statistical uncertainty estimates}\label{appendix:A}

We looked at  23 runs of the RG-C collected data on a polarized $ND_3$ target. The data corresponded to $eD \rightarrow e'\pi^{+}X$, and due to the low tensor polarization, we looked at the events in a wide kinematic range: $Q^2 > 0.95$~GeV$^2$, $0 < P_T < 0.8$~GeV, $0.08 < x_{B} < 0.8$ and $0.2 < z < 0.7$.

A crude estimate was made by assuming that the total cross-section was given by 

\begin{equation}
    \sigma=\sigma_{u}^D+S_{\parallel}\sigma _{S}+T_{\parallel\parallel}\sigma _{T} + \sum _{i} \sigma ^{i}
    \label{eq:sigma_all}
\end{equation}

\noindent where $\sigma_{S} = 0.3\sigma_{u}^D$ and $\sigma_{T} = 0.1\sigma_{u}^D$, here $\sigma_{u}^D$ 
is the unpolarized cross-section of deuteron, $\sigma_{S}$ is the vector polarized cross-section, and $\sigma_{T}$ is the tensor polarized cross-section. In general, it is the sum of the deuteron cross-section and the cross-section of the nuclei in the sample, e.g. i = Nitrogen (N), Helium (He) etc. The 0.3 and 0.1 factors are estimates regarding the vector and tensor contributions compared to the unpolarized cross section, and all add to the total cross section. The dilution factor \(f\) was taken as 0.27, and the vector and tensor polarizations were taken from the estimates in Fig.~\ref{fig:nd3_pol_vecten}. We observed that the number of events corresponding to $eD \rightarrow e'\pi^{+}X$ was consistently 1.7\% of the total events in all runs. 

To estimate the expected uncertainty, we evaluated two cases. First assuming that the negative and positive vector polarizations are the same, and second accounting by the difference in polarization.

\subsection{$S^+_{\parallel} = -S^-_{\parallel}$ Approximation}

In the scenario that $S^+_{\parallel} =- S^-_{\parallel} = S_{\parallel}$, what we measure is:
\begin{equation}
\begin{split}
\sigma _{meas} ^{+}  = \frac{1}{2} \bigg( \sigma _u ^D + S_{\parallel}^{+} \sigma _S   + T_{\parallel \parallel}\sigma _T + \sum _{i} \sigma ^{i}  \bigg)\\
\sigma _{meas} ^{-}  = \frac{1}{2} \bigg( \sigma _u ^D + S_{\parallel}^{-} \sigma _S   + T_{\parallel \parallel}\sigma _T + \sum _{i} \sigma ^{i} \bigg)  \\
\sigma _{meas} ^{total} = \sigma _{meas} ^{+}  + \sigma _{meas} ^{-} =  \sigma _u ^D + T_{\parallel \parallel}\sigma _T + \sum _{i} \sigma ^{i} 
\end{split}
\label{eq:4}
\end{equation}
the factor of $\frac{1}{2}$ assumes that the data is equally divided between positive and negative polarization.  

Therefore, by subtracting the unpolarized cross-section $\sigma_{meas}^u (= \sigma_u^D+\sum _{i} \sigma ^{i})$ and rearranging the terms, we can obtain:

\begin{equation}
   T_{\parallel \parallel}\frac{\sigma _T}{\sigma _u ^D} = \frac{1}{f} 
   \frac{\sigma _{meas} ^{total} - \sigma _{meas} ^{u}}{\sigma _{meas} ^{u}}
\label{eq:8}
\end{equation}

We are interested in finding the error propagation of $T_{\parallel \parallel}\frac{\sigma _T}{\sigma _u ^D}$, which will be re-label $R_d^T$, for easier reading. The statistical uncertainty is given by,

\begin{equation}
    (\delta R_d^T) ^{stat} = \frac{1}{f} \sqrt{\frac{ (\sigma _{meas} ^{u})^2 N_{meas} + (\sigma _{meas} ^{total})^2 N_{meas} ^{u}}{(\sigma _{meas} ^{u})^4}}
\label{eq:10}
\end{equation}

If $\delta _{\sigma _{meas} ^{total}} ^{stat} = \sqrt{N_{meas}}$  and $\delta _{\sigma _{meas} ^{u}} = \sqrt{N_{meas} ^{u}}$, and all acceptance and efficiencies can cancel in the ratio:

\begin{equation}
\frac{\sigma _{meas} ^{total} - \sigma _{meas} ^{u}}{\sigma _{meas} ^{u}} = \frac{N _{meas}  - K_1 N _{meas} ^{u}}{K_1 N _{meas} ^{u}} 
\label{eq:11}
\end{equation} 

where $K_1 = \frac{Q _{meas}}{Q _{meas}^u}$ is the difference in the accumulated charge for both measurements. 

Then, Eq.~\ref{eq:10} changes to:
\begin{equation}
    (\delta R_d^T) ^{stat}  =  \frac{1}{f} \sqrt{\frac{(K_1 N _{meas} ^{u})^2 N_{meas} + (N _{meas})^2 N_{meas} ^{u}}{(K_1 N _{meas} ^{u})^4}}
\label{eq:12}
\end{equation}

\begin{figure}[htb]
    \centering
    \includegraphics[height=0.65\linewidth]{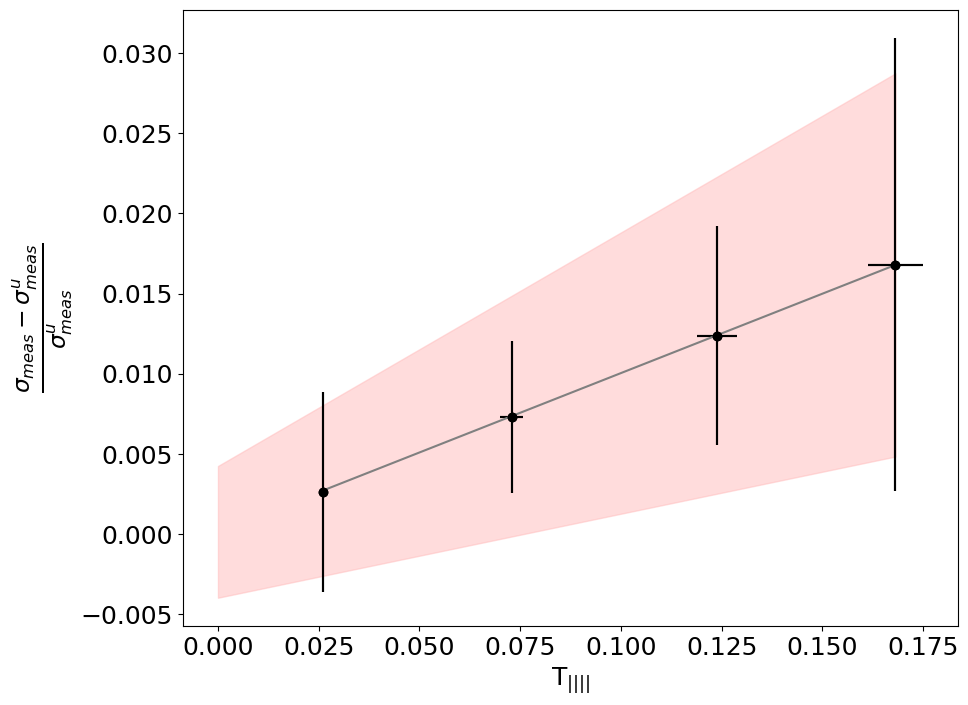}
    \caption{Measured quantity with respect to the tensor polarization. The error bars were calculated using Eq.~\ref{eq:12}, the fit with the uncertainty band was conducted using generated pseudo-data consistent with the expected uncertainty measurements.}
    \label{fig:t1} 
\end{figure}

Using Eq.~\ref{eq:12} after grouping the runs in tensor polarization average between ranges of 0.05, we get Fig.~\ref{fig:t1}. 

By generating pseudo-data within the error bars including the 7\% relative uncertainty of the dilution factor, as shown in Fig.~\ref{fig:t1}. The expected measured slope is $0.099 \pm 0.046$. 

To extract the structure function, we still have a few steps to follow. First, the unpolarized and polarized cross-sections can be approximated as: 

\begin{equation}
    \sigma _u  = \frac{y^2\alpha^2}{2(1-\epsilon)xyQ^2} \left(1+\frac{\gamma^2}{2x}\right) F_{UU,T}
    \label{eq:13}
\end{equation}

\begin{equation}
    \sigma _T  = \frac{y^2\alpha^2}{2(1-\epsilon)xyQ^2} \left(1+\frac{\gamma^2}{2x}\right) F_{U(LL),T}
    \label{eq:14}
\end{equation}

Therefore,
\begin{equation}
    \frac{\sigma _T}{\sigma _u}  = \frac{F_{U(LL),T}}{F_{UU,T}}
    \label{eq:15}
\end{equation}

The uncertainty will be dominated by statistics, which we predict, under this assumption, will be approximately$\sim$46\%.

\subsection{$S^+_{\parallel} \neq -S^-_{\parallel}$ as in the experiment}

More realistically the vector polarizations are not the same, 
\begin{equation}
\begin{split}
\sigma _{meas} ^{+}  =  \sigma _u ^D + S_{\parallel}^{+} \sigma _S   + T_{\parallel \parallel}^{+} \sigma _T + \sum _{i} \sigma ^{i}  \\
\sigma _{meas} ^{-}  =  \sigma _u ^D + S_{\parallel}^{-} \sigma _S   + T_{\parallel \parallel}^{-} \sigma _T + \sum _{i} \sigma ^{i}   
\end{split}
\label{eq:4}
\end{equation}
therefore, to cancel the vector cross-section contribution:
\begin{equation}
S_{\parallel}^{-} \sigma _{meas} ^{+}  + S_{\parallel}^{+} \sigma _{meas} ^{-} = (S_{\parallel}^{-}  + S_{\parallel}^{+})\sigma _u ^D + (S_{\parallel}^{-} T_{\parallel \parallel}^{+}  + S_{\parallel}^{+} T_{\parallel \parallel}^{-})\sigma _T  + (S_{\parallel}^{-}  + S_{\parallel}^{+})\sum _{i} \sigma ^{i} 
\end{equation}

We can rearrange the last equation as follows:

\begin{equation}
\frac{(S_{\parallel}^{-} T_{\parallel \parallel}^{+}  + S_{\parallel}^{+} T_{\parallel \parallel}^{-}) }{(S_{\parallel}^{-}  + S_{\parallel}^{+})} \frac{\sigma _T}{\sigma _u ^D } = \frac{1}{f} \frac{S_{\parallel}^{-} \sigma _{meas} ^{+}  + S_{\parallel}^{+} \sigma _{meas} ^{-} - (S_{\parallel}^{-}  + S_{\parallel}^{+})\sigma _u}{(S_{\parallel}^{-}  + S_{\parallel}^{+})\sigma _u}
\end{equation}

The statistical uncertainty of the measured quantities are: $\delta _{\sigma _{meas} ^{+}} ^{stat} = \sqrt{N_{meas}^{+}}$, $\delta _{\sigma _{meas} ^{-}} ^{stat} = \sqrt{N_{meas}^{-}}$, $\delta _{\sigma _{u}} ^{stat} = \sqrt{N_{u}}$. The uncertainty of the measured polarizations on the right side of the equation is currently neglected, assuming that by binning the data in terms of polarization and taking the ratio, these uncertainties will be suppressed.

Renaming, $(R_d^T) ^{stat} =\frac{(S_{\parallel}^{-} T_{\parallel \parallel}^{+}  + S_{\parallel}^{+} T_{\parallel \parallel}^{-}) }{(S_{\parallel}^{-}  + S_{\parallel}^{+})} \frac{\sigma _T}{\sigma _u ^D}$, 

\begin{equation}
\Scale[0.9]{
(\delta R_d^T) ^{stat}= \frac{1}{f}\sqrt{\frac{(N _u)^2  ((S_{\parallel}^{-}w^+ \delta ^{N _{meas} ^{+}})^2 + (S_{\parallel}^{+}w^- \delta ^{N _{meas} ^{-}})^2) + (\delta ^{N _{u}})^2( S_{\parallel}^{-} w^+ N _{meas} ^{+}  + S_{\parallel}^{+} w^- N_{meas} ^{-})^2 }{(S_{\parallel}^{-}  + S_{\parallel}^{+})^2(N _u)^4 }}
},
\label{eq:meth2}
\end{equation}
where $w^+ = Q^u/Q^+$ and $w^- = Q^u/Q^-$. We assume that all acceptance, efficiency, and other corrections can be canceled out in the ratio.

\begin{figure}[htb]
    \centering
\includegraphics[height=0.65\linewidth]{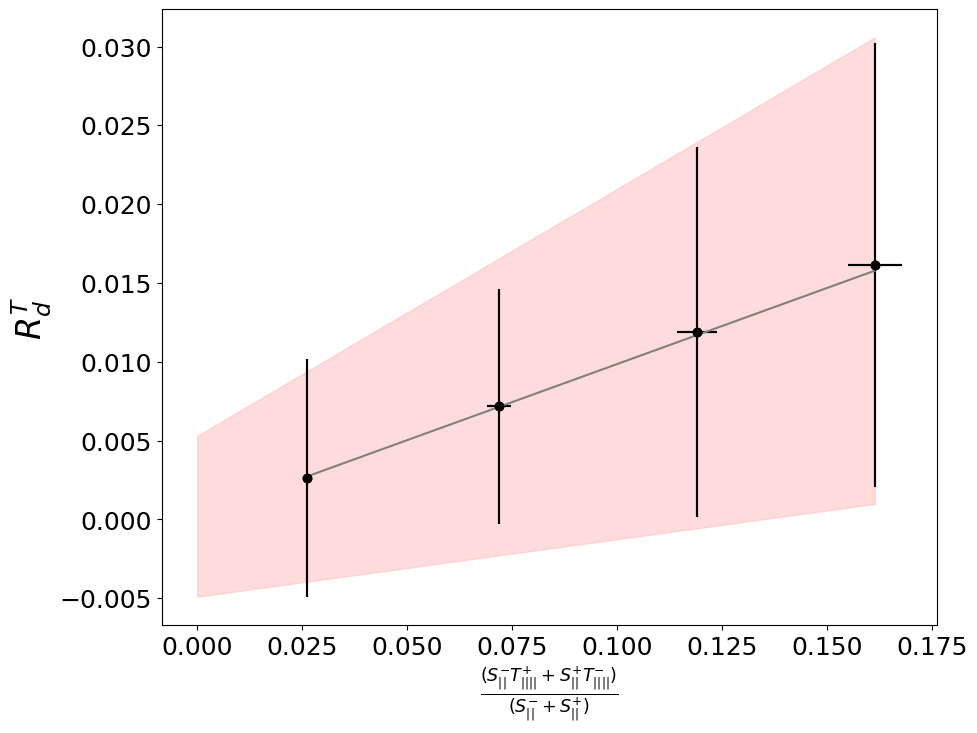}
    \caption{Expected measurements accounting for differences in positive and negative polarization. Uncertainties are evaluated using Eq.~\ref{eq:meth2}, the fit with the uncertainty band was conducted using generated pseudo-data consistent with the expected uncertainty measurements.}
    \label{fig:pseudo-data2} 
\end{figure}

Using this method, the pseudo-data generated is shown in Fig.~\ref{fig:pseudo-data2}. The expected measured slope is $0.103 \pm 0.061$. Then, a 60\% uncertainty will be expected.

This first-ever extraction will be crucial for our planned exploratory measurements of these quantities. Though a 60\% error bar would be considered poor precision for a dedicated measurement, it is more than sufficient from this analysis to provide an initial guide for the rates and kinematics needed to measure the structure functions with high precision.

\end{document}